\documentclass{emulateapj}
\usepackage{latexsym}
\usepackage{graphicx}
\usepackage{amssymb}
\usepackage{enumerate}
\usepackage{ulem}
\usepackage[usenames]{color}

\shorttitle{X-ray emission and dynamics from superbubbles}

\begin{document}

\title{X-ray emission and dynamics from large diameter superbubbles: The case
of N\,70 superbubble}

\author{Rodr\'iguez-Gonz\'alez, A.\altaffilmark{1,2}, Vel\'azquez, P. F.\altaffilmark{1},
  Rosado M.\altaffilmark{2},  Esquivel, A.\altaffilmark{1},Reyes-Iturbide, J.\altaffilmark{3} \& Toledo-Roy, J. C.\altaffilmark{1}}
\affil{$^1$ Instituto de Ciencias Nucleares, 
Universidad Nacional Aut\'onoma de M\'exico, 
Apartado Postal 70-543, 04510 D.F., M\'exico\\
$^2$ Instituto de Astronom\'\i a,
Universidad Nacional Aut\'onoma de M\'exico,
Apartado Postal 70-264, 04510 D.F., M\'exico\\
$^3$ Escuela Superior de F\'\i sica y Matem\'aticas, IPN, 
U.P. Adolfo L\'opez Mateos, C.P. 07738 D.F., M\'exico.}
\email{ary@nucleares.unam.mx}




\begin{abstract}
The morphology, dynamics and thermal X-ray emission of the superbubble
N70 is studied by means of 3D hydrodynamical simulations, carried
out with the {\sc{yguaz\'u-a}} code. We have considered different
scenarios: the superbubble being the product of a single supernova
remnant, of the stellar winds from an OB association, or the result of 
the joint action of stellar winds and a supernova event. Our results
show that, in spite that all scenarios produce bubbles with the
observed physical size, only those where the bubble is driven by
stellar winds and a SN event are successful to explain the general
morphology, dynamics and the X-ray luminosity of N70. Our models
predict temperatures in excess of $10^8~\mathrm{K}$ at the interior of
the superbubble, however the density is too low and the emission in
thermal X-ray above $2~\mathrm{keV}$ is too faint to be detected.
\end{abstract}

\section{Introduction}

Massive OB stars, or groups of them (young clusters or OB
associations) inject  a large amount of mechanical energy via stellar
winds and violent supernova (SN) episodes to the interstellar medium
(ISM).
They sweep-up their environment producing the so-called bubbles and
superbubbles (when they are produced by a single star, or multiple
stars, respectively).
The standard models of these bubbles are those by Weaver et
al. (1977), and Chu \& Mac Low (1990). They consider the mechanical
energy input of a stellar wind, and predict an extended bubble
structure of shock-heated gas that emits mainly in X-rays, surrounded
by a cool shell of swept-up material that is bright at optical
wavelengths.
These models has been compared with several observations, and
the X-ray observed luminosities often exceed the theoretical
predictions (i.e. Chu \& Mac Low 1990, Wang \& Helfand 1991).

Later, Oey (1996b) based on the observations of Rosado et
al. (1981, 1982), and Rosado (1986) proposed two categories of
superbubbles: high-velocity and low-velocity ones.
The high-velocity superbubbles are characterized by a shell expansion
velocity $v_s \gtrsim 25~\mathrm{km~s^{-1}}$, and they are as common as the
low-velocity ones (e.g. Rosado 1986).
The difference, however, lies in the fact that it is virtually
impossible to obtain expansion shell velocities in excess of
$25~\mathrm{km~s^{-1}}$ in superbubbles with large diameters (about 100~pc) 
without additional acceleration (e.g. an impact from a supernova
remnant, SNR). 
The energy injected by SN explosions would be an extra source of
heating for the gas inside the superbubble and this could explain the
observed X-ray excess.

In this work we have turned our attention to the superbubble N\,70 in
the Large  Magellanic Cloud (LMC). 
N\,70 is an almost circular superbubble of approximately 50\,pc in radius. 
The superbubble is driven by the OB association LH\,114 
(Lucke \& Hodge 1970), which contains more than a thousand stars.
Oey (1996a) classified seven of them as a O-type stars and estimated
the mean age of the OB association to be around 5~Myr.
Rosado et al.(1981) and Georgelin et al.~(1983) found 
[SII]/H$\alpha$ line-ratios in N\,70 with values larger than 
those in photoionized \ion{H}{2} regions, but lower than those of SNR in the
LMC.  The measured expansion velocity of this superbubble ($\sim
70~\mathrm{km~s^{-1}}$) is consistent with shock models that also
reproduce the  [\ion{S}{2}]/H$\alpha$ ratio of Rosado et al. (1981). 
However, the dynamical age derived with this velocity  does not agree
with the model of  Oey (1996b).

Reyes-Iturbide et al. (2011) calculated the
thermal X-ray luminosity for the superbubble N\,70 (DEM\,301), with the
XMM-Newton observations from Jansen et al. (2001). For the analysis of
the  X-ray spectrum they used the three individual data sets adjusting
them jointly. They extracted spectra from each of the three EPIC/MOS1,
EPIC/MOS2 and EPIC/PN event files.
The spectra were fitted with a two-component model consisting of a
thermal plasma-MEKAL (Kaastra \& Mewe 1993), and nonthermal power law. The 
resulting spectra were analyzed jointly using the XSPEC spectral
fitting package, where the fit has an absorption column density of 
$N_H=1.4\pm 0.5 \times 10^{20}$~cm$^{-2}$
(in agreement with the measures of column densities in the LMC direction, 
see Dickey \& Lockman 1990). The X-ray luminosity in the 0.2-2~keV energy
band with absorption-corrected was found to be
1.6$\times$10$^{35}$~erg~s$^{-1}$.

Here, we present a series of 3D numerical simulations of the N\,70 
superbubble using the physical properties (stellar types, positions, 
etc.) of the stellar cluster in its interior.
We analyze the resulting morphology, dynamics and thermal X-ray
emission, and compare it with observations of N\,70.
The paper is organized as follows. In Section~2 we provide a brief
review of the models and theoretical predictions of the emission in
superbubbles.  In Section~3 we describe the numerical simulations, the
results of the simulations are analyzed in Section~4, and a summary
is provided in Section~5.

\section{Superbubble dynamics and X-ray emission}

\begin{table*}
\begin{center}
\caption{Coordinates and spectral types of the most massive stars 
inside N\,70}
\begin{tabular}{lccccc}
\tableline\tableline
Star & RA & DEC & spectral type & $V_{\infty}$ &$\log\left( \dot{M}\right)$\\
~ & [hr min sec] & $[\degr~\arcmin~\arcsec]$ & ~& $[\mathrm{km~s{^-1}}]$ &
   $[\mathrm{M}_{\odot}~\mathrm{yr}^{-1}]$\\  

\tableline
D301-1005 &5 43 08.33 &-67 50 52.5 &O9.5 V &1500 &-6.9\\
D301SW-1a &5 43 15.50 &-67 51 09.7 &O8 III(f) &2000&-6.6\\
D301SW-1b &5 43 15 50& -67 51 09.7 &O9: V &1500 &-6.8\\
D301SW-3 & 5 43 12.87 &-67 51 16.3 &O3 If &4100 &-4.90 \\
D301NW-4 &5 43 17.70 &-67 50 36.6  &O5: III:e &2900 &-6.2 \\
D301NW-8&5 43 15.98&-67 49 51.2&O7 V((f))&2000&-6.6 \\
D301NW-9& 5 43 24.60& -67 50 31.1&O9.5 V&1500 &-6.9 \\
D301NE-5& 5 43 34.85& -67 50 40.9&B0.5 V&2000&-7.25 \\
D301NW-12& 5 43 23 79& -67 50 21.5&BO V&2000&-7.3 \\
D301NW-13& 5 43 06.71& -67 49 56.0&B1 V&1700&-6.51 \\
D301SW-9&5 43 10.03& -67 52 21.3&B1.5: V&900&-5.26 \\
D301NW-15& 5 43 12.25& -67 50 52.8&B1.5V&900&-5.26 \\
D301NW-18& 5 43 11.13& -67 50 40.3& B0 V&2000&-7.3\\
\tableline
\end{tabular}
\end{center}

\end{table*}

Let us consider a simple model of superbubble formation, where the stars 
deposit the total mechanical energy in form of stellar winds. Such
mechanical luminosity is given by 
\begin{equation}
\label{eq1}
L_w=\sum^N_{i=1}\frac{1}{2}\dot{M}_{w,i} v^2_{w,i}\, ,
\end{equation}
where, $\dot{M}_{w,i}$ and $v_{w,i}$ are the mass-loss rate and 
the wind terminal velocity of the $i$-th star, and $N$ is the total
number of stars.
At the beginning, the stellar winds inside the cluster volume collide
with the surrounding  ISM (here we will assume a uniform medium with
preshock number density ``$n_0$'') forming shells of shocked ISM
material. At some point the volume between the stars fills with
shocked material from the individual stars and the winds coalesce into
a common cluster wind that forms a larger shell, a ``supershell''
(Cant\'o et al. 2000, Rodriguez-Gonzalez et al. 2008, etc.). 
As the supershell expands with respect to the cluster center, one can
distinguish a  superbubble structure with four regions:
\begin{enumerate}[a)]
\item A free wind region, formed by unperturbed stellar wind,
which is only found around the most powerful stars.
\item A shocked wind region, formed by the interaction of several
  individual stellar winds. This material has been heated enough that
  it emits primarily in X-rays.
\item An outer region of swept-up interstellar medium with an important
  optical line emision.
\item The unperturbed ISM medium (of uniform density of $n_0$), just
  outside the swept-up shell.
\end{enumerate}
 
The X-ray luminosity that arise from the internal shocked region,
where the gas temperature is in the range of 10$^6$-10$^7$~K 
($\sim0.1-2$~keV), can be estimated as in Weaver et al. (1977) and Chu \& Mac
Low (1990)~:
\begin{equation}
\label{lx}
L_X=3.29 \times 10^{34}I(\tau) \xi L^{33/35}_{37} n^{17/35}_0
t^{19/35}_6 \,\,\, [{\rm erg\, s^{-1}}] 
\end{equation}
where
\begin{equation}
  I(\tau)=\frac{125}{33}-5\tau^{1/2}+\frac{5}{3}\tau^3-\frac{5}{11}\tau^{11/3},~\mathrm{and}
\end{equation}
\begin{equation}
  \tau=0.16L^{-8/35}_{37}n^{-2/35}_0 t^{6/35}_6.
\end{equation}
$\xi$ is the gas metallicity, $L_{37}=L_w/10^{37}$, $t_6=t/10^6$,
$L_w$ is the mechanical luminosity of the cluster, and $t$ is the
cluster lifetime.  If a supernova explodes at the center of stellar
cluster, the total X-ray luminosity will be modifed as estimated by
Chu \& Mac Low (1990)~:
\begin{displaymath}
L(SNc)_X=8\times 10^{33} \xi h(x_s)(1-x_s)^{-2/5} L^{33/35}_{37} \\
\end{displaymath}
\begin{equation}
\label{lxsn}
\;\;\;\,\,\,\,\,n^{17/35}_0 t^{19/35}_6\,\,\,[{\rm erg \,s^{-1}}]
\end{equation}
where, $x_s=r_s/R$, $r_s$ is the radius of the remnant,
$R$ is the radius of the superbubble, and
\[
h(x)=\frac{125}{156}-\frac{5}{13}(1-x)^{13/5}+\frac{5}{4}(1-x)^{8/5}\\
\]
\begin{equation}
-\frac{5}{3}(1-x)^{3/5}.
\end{equation}
However, as mentioned above, observed X-ray luminosities exceed these
predictions. In order to explain such differences several
alternatives have been explored, for instance: Chu \& Mac Low
(1990) have proposed an off-centered supernova explosion, Silich et
al. (2001) studied effects of metallicity enhancement (due to
evaporation of the  outer shell) and Reyes-Iturbide et al. (2009)
considered the interaction of the cluster wind with a high density
region in the ISM for the case of M\,17.

For N\,70, the total mechanical luminosity injected by the stellar winds
of massive stars (the most massive are listed in Table~1) is around 
7.31$\times$10$^{37}$~erg s$^{-1}$. This superbubble evolves in an ISM
with number density $\sim$0.16~cm$^{-3}$ (Rosado et al. 1981 and Skelton et 
al. 1999), and an average gas metallicity is $\sim 0.3
\mathrm{Z}_\odot$ (typical of the LMC, Rolleston, Trundle \& Dufton 2002). 
N\,70 is quite circular with a  
radius of $\sim$50~pc, using the shell expansion velocity a dynamical 
age of $\sim$3$\times$10$^5$~yr can be obtained. Using these values
in the equations~\ref{lx} and \ref{lxsn}, the predicted
X-ray luminosity for this object is  3.32$\times$10$^{34}$~erg
s$^{-1}$ when only the stellar winds are taken
into  account, and 3.68$\times$10$^{34}$ erg s$^{-1}$ if one adds a
single centered SN to the cluster wind.

The X-ray luminosities predicted by the standard models are an order
of magnitude less than the observed value. The difference seems to
large to be explained by the metallicity effects as proposed by Silich
et al. (2001), and the ISM around it is fairly homogeneous (unlike in
M\,17 where the inhomogeneity of the medium suffices to explain the
X-ray luminosity). 
In addition, Oey (1996b) showed that is essentially impossible 
to obtain expansion velocities ($\gtrsim$25~km s$^{-1}$) in superbubbles 
with radius of a few tens of parsecs without induced acceleration 
(a SNR impact was proposed in that paper). Thus, given the high X-ray
luminosity and expansion velocity we chose to consider an off-centered SN
explosion, with the restriction  that it can not be too far from the
center because the quasi-spherical shape of N\,70. 
The SN possibility is also consistent with the stellar population
models N\,70 presented by Oey (1996b) where 13 massive stars 
are found the range form 12 to 40~M$_\odot$). A $\sim$60~M$_{\odot}$ star
could be expected using a standard initial mass function of N\,70, and
if  formed with the rest of the cluster, it would already have
exploded as a SN. 

Table~1 shows the coordinates and spectral types of the most massive
stars  inside N\,70. In the same table, we have included
characteristic values of the terminal wind speed and mass loss rate
associated with stars of such spectral types (de Jager et al. 1988, Wilson \& 
Dopita 1985, Leitherer 1988, Prinja et al. 1990, Lamers \& Leitherer 1993, 
Fullerton et al. 2006).

\section{The numerical models}

\begin{figure}
\includegraphics[width=7.5cm]{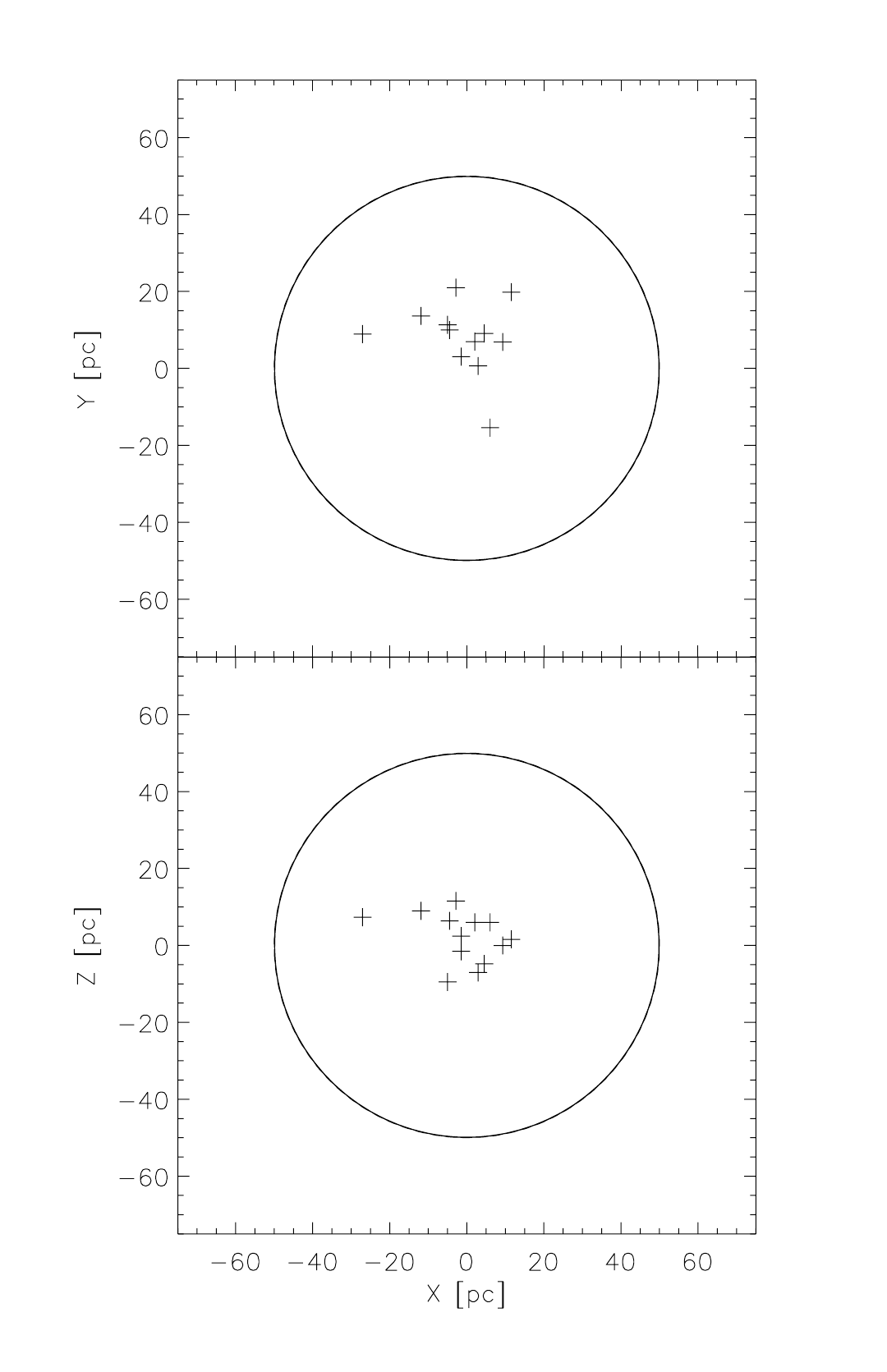}
\caption{The stellar distribution of $xy$-plane (top panel) and x-z plane 
(bottom panel) for all the numerical models.}
\end{figure}

In order to estimate the X-ray emission and shell dynamics in N\,70, we 
have computed 3D numerical simulations with the full, radiative
gas dynamic equations. We use  a tabulated cooling function obtained
with the CHIANTI~\footnote{The CHIANTI database and associated IDL 
procedures, now  distributed as version 5.1, are freely available at:  
http://wwwsolar.nrl.navy.mil/chianti.html and
http://www.arcetri.astro.it/science/chianti/chianti.html}  database,
using a metallicity  $\xi$=0.3\,Z$_\odot$ (consistent with that the
LMC, see Rolleston, Trundle \& Dufton 2002).  
The simulations include multiple stellar wind sources in the 3D
adaptive grid {\sc yguaz\'u-a} code, 
which is described in detail by Raga et al. (2000, 2002). They were 
computed with a maximum resolution of 0.4296~pc (corresponding to $256^3$ 
grid points at the maximum grid resolution) in a computational domain of 
110 pc (along each of the 3 coordinate axis). We have not included
thermal conduction effects in any of our models.

In all runs, we assumed that the computational domain was initially
filled by a homogeneous ambient medium with temperature
$T_{0}=$ 10$^4$~K (as it would be expected in the photoionized region around 
the massive OB association) and density $n_{0}=0.16$~cm$^{-3}$. The stellar 
winds
are imposed in spheres of radius $R_w=7.94\times 10^{18}$~cm ($\sim$0.58~pc), 
corresponding to 6~pixels of the grid. 
 Table 1 gives the position of the stars in equatorial coordinates (J2000), 
which can be translated to parsecs considering that the cluster is at a 
distance of 50~kpc.
Then the wind sources are placed in the $xy$-plane 
according to their positions in the sky. Since we do not know the individual 
line-of-sight distance ($z$-coordinate) to the stars, we produced randomly 
picked positions in $z$, retaining the same $xy$ configuration.
The $z-$distribution was obtained from a pseudo random sampling to yield a 
$\propto R^{-2}$ distribution (similarly to Reyes-Iturbide at al 2009).
 The maximum of the distribution from which the z positions were sampled was 
set to the maximum separation in the plane of the sky.
Figure~1 shows the stellar distribution in the $xy$-plane (top panel)
and $xz$-plane (bottom panel) for all the numerical models.
Inside the spheres centered at the star positions a stationary wind is
imposed (at all times) with  an $\propto R^{-2}$ density
profile scaled to yield the $V_{\infty}$ and $\dot{M}$ for each star, and 
a constant temperature $\propto V^2_{\infty}$.

We ran four numerical models, M1, M2, M3 and M4 to explore the efects of
the mechanical energy injected by the stellar winds and the 
supernova explosion in the superbubbles dynamics and X-ray
emission. The properties of the models are presented in Table~2.
In model M1, we considered the energy injected by a single SN
(with 10$^{51}$~erg) in an homogeneous ISM. In model M2 included
the mechanical energy injected by  the stellar winds alone. 
Models M3 and M4 explore the combined effect of stellar winds and a SN
explosion. For model M3 we included the energy injected by a SN
(similar to that in M1) inside the wind blown bubble (as model M2) at the
center of the stellar population of N\,70. The SN detonation was
imposed at t=1.15$\times$10$^5$~yr.  Finally, in model M4 we explored
the effects of a SN slightly off-center, the SN 
explosion was placed at (1.5, -1.5, -1.5)~pc from  the center of the
stellar distribution, also at t=1.15$\times$10$^5$~yr.

\begin{table}
\begin{center}
\caption{Numerical models general properties}
\begin{tabular}{cccccr}
\tableline\tableline
 Model      &   Winds     &   SN     &   SN Location \\
\tableline
M1          &    no       &   yes    &     Center\\
M2          &  13 stars   &   no     &      no\\
M3          &  13 stars   &   yes    &     Center\\
M4          &  13 stars   &   yes    &     Off-center\\
\tableline
\end{tabular}
\end{center}
\end{table}

\section{Results}
\subsection{Superbubble dynamics}

In order to obtain the physical flow configuration  we computed 
 the radially dependent flow density, radial velocity and temperature
averaging over spherical concentric surfaces $S_R=4 \pi R^2$~(see also
Rodr\'iguez-Gonz\'alez et al. 2007):

\begin{equation}
\rho_a(R)={1\over 4\pi}{\int_{S_R} \rho\,\sin\theta\,d\theta d\phi}\,,
\label{ra}
\end{equation}

\begin{figure}
\includegraphics[width=7.5cm]{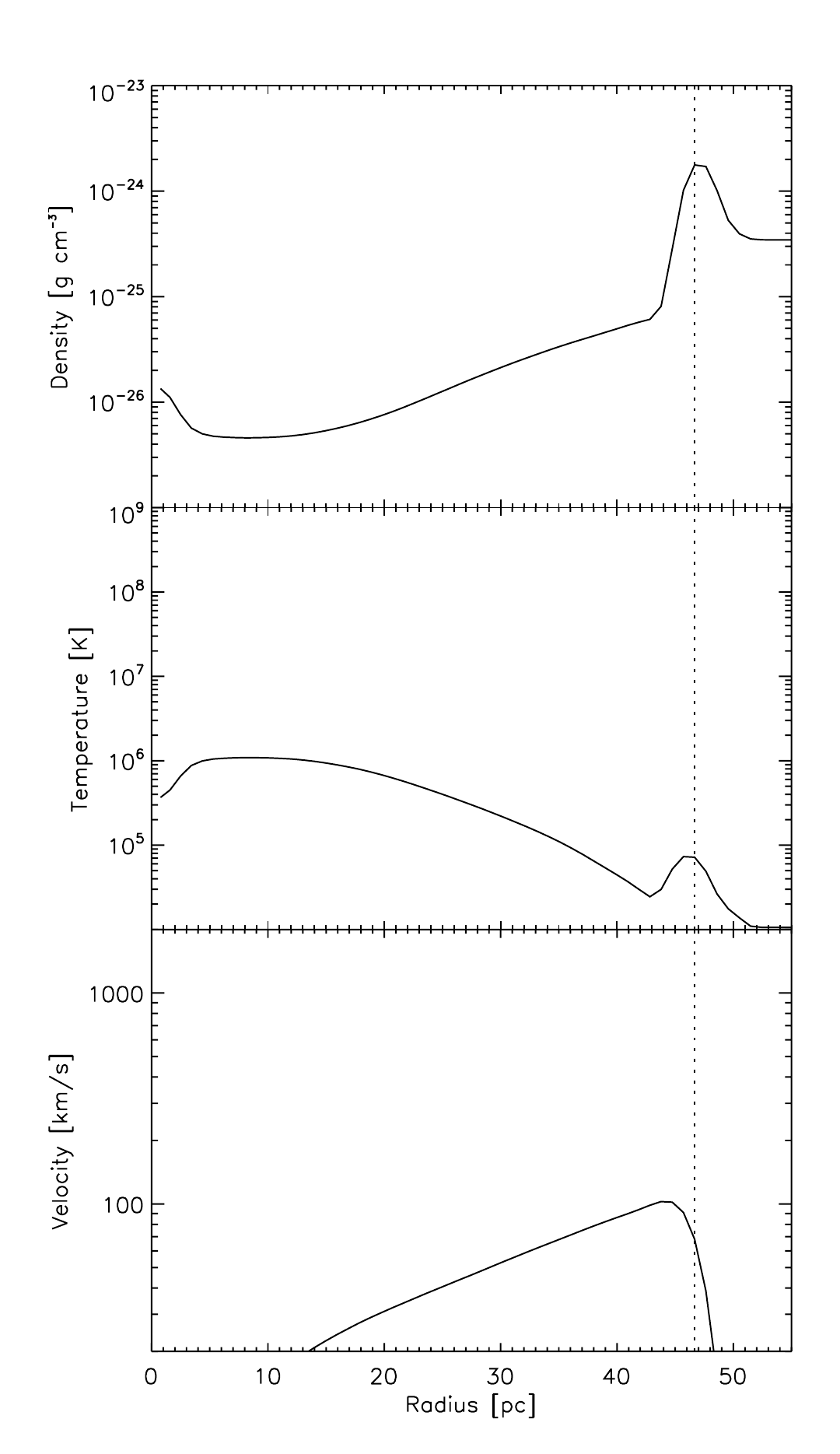}
\caption{The spherically averaged flow from model M1. the density (top),
temperature (center) and radial velocity (bottom) obtained from the 
numerical simulation is shown as a function of spherical radius $R$. 
The dashed lines represent the position of the maximum value of the shell 
density.}
\end{figure}

\begin{equation}
v(R)={1\over 4\pi \rho_a(R)}{\int_{S_R} \rho v_R\,
\sin\theta\,d\theta d\phi}\,,
\label{va}
\end{equation}
\begin{equation}
T(R)={1\over 4\pi \rho_a(R)}{\int_{S_R} \rho T\,
\sin\theta\,d\theta d\phi}\,,
\label{ta}
\end{equation}
where $\theta$ and $\phi$ are the polar and azimuthal angles, respectively, 
$\rho$ is the flow density, $T$ the temperature and
$v_R$ the radial velocity (obtained by projecting the
three cartesian velocity components resulting from the numerical
integration onto the direction normal to the spherical surface). That 
is $v_R=(x v_x+ y v_y+ z v_z)/R$.

Figures~2, 3 and 5  show the superbubble and shell distributions of
density, temperature and radial velocity (top, middle and bottom 
panel) for model M1, M2 and M3, respectively, at an evolutionary time
of 2$\times$10$^5$~yr.  
Model M1 (see Figure~2) forms a thin shell with maximum 
density at R=47~pc. This shell contains the interstellar medium that
has been swept up by the leading shock produced by the explosion. The
gas behind the leading shock cools and forms the thin
shell. There the temperature is around 10$^5$~K, in the
range of optical line emission.  At the radius at which the density is
maximum the radial velocity is around 75~km~s$^{-1}$.  This  model does not 
include the stellar wind contribution and the radial velocity drops
because of the interior of the bubble is cooling radiatively (the supernova
remnant has past the Sedov phase and it is well into the radiative one).

\begin{figure}
\includegraphics[width=7.5cm]{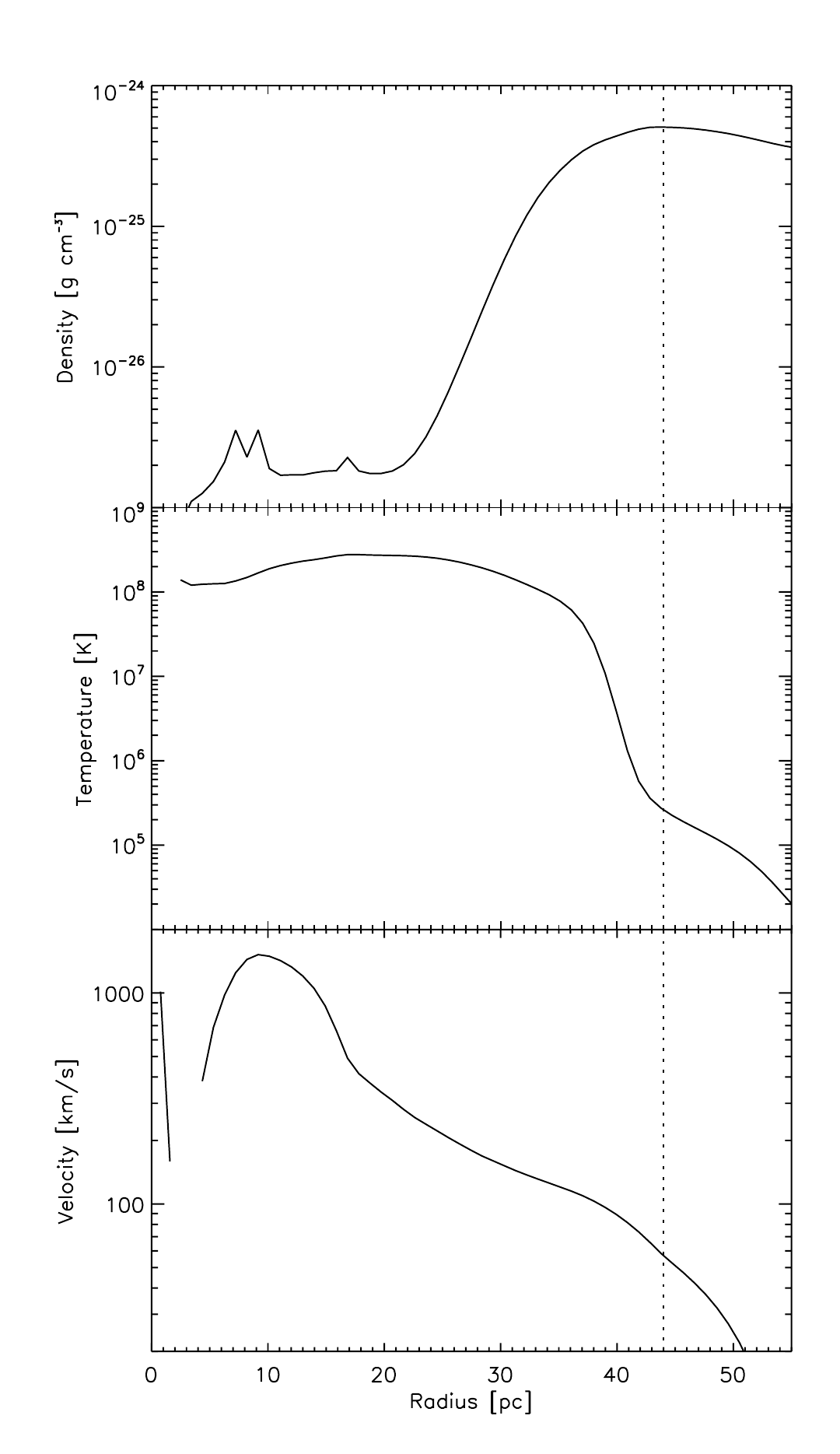}
\caption{Same as Figure~2 but for the  model M2.}
\end{figure}

The contribution of the stellar winds of the cluster in the shell 
dynamics is present in Figure~3. Model M2 (see Figure~3 and 4) presents a 
thick shell  with a maximum density at R=44~pc. This shell is driven by the 
mechanical energy injected by stellar winds inside the cluster volume
in the form of a common cluster wind (Cant\'o et al. 2000, 
Rodr\'iguez-Gonz\'alez et al. 2008, etc.).

\begin{figure}
\begin{center}
\includegraphics[width=7.5cm,angle=-90]{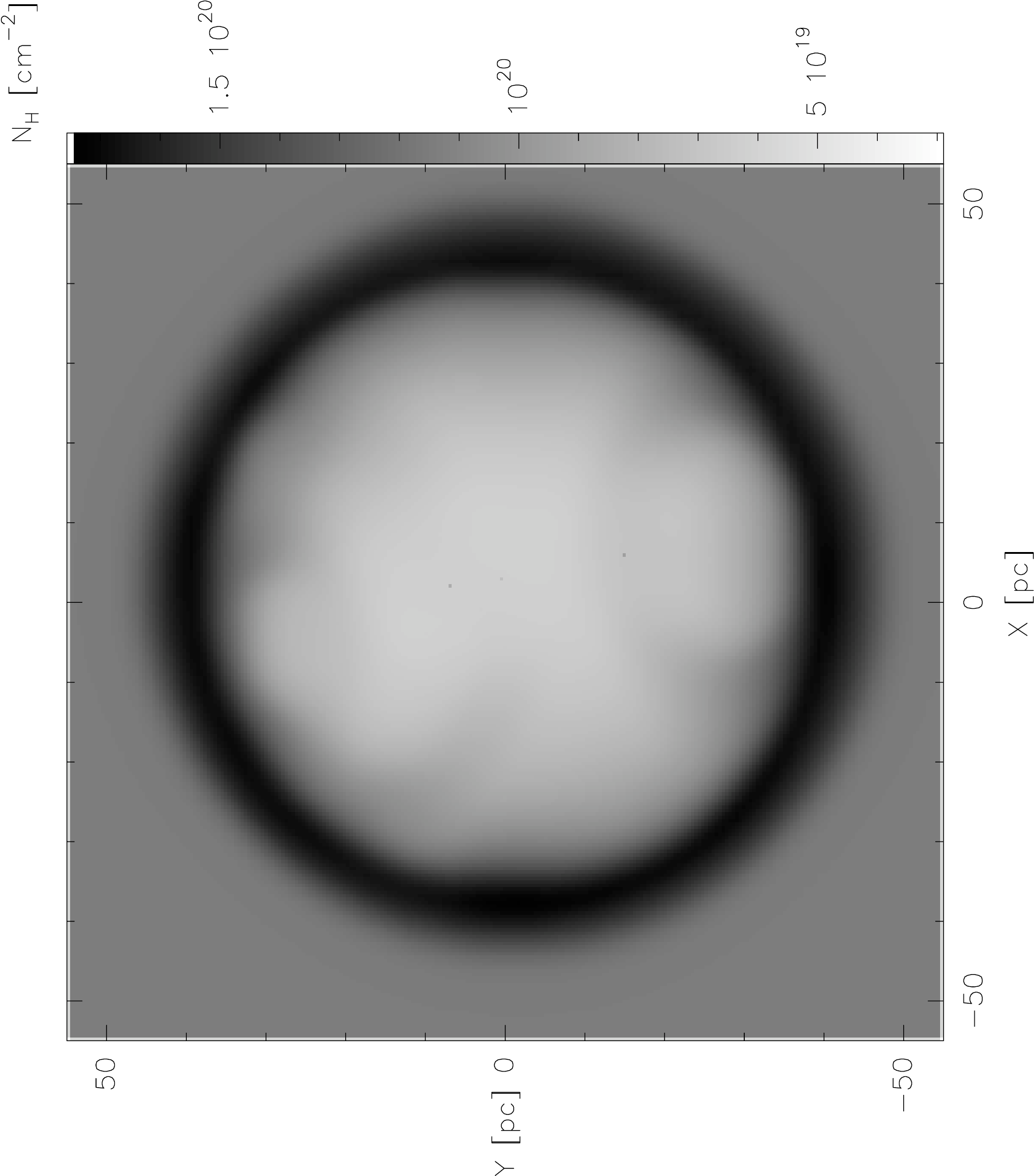}
\caption{The column density for model M2 at
  t=2$\times$10$^5$~yr.}
\end{center}
\end{figure}

Figure~3 shows a average temperature (inside the shell) of 
5$\times$10$^5$~K (optical line emission regime) and the radial velocity 
at the density peak is around 45~km~s$^{-1}$ as predicted by the
standard model of  Weaver et al. 1977 (see also Chu et  al. 1995).
This velocity is, however, lower than that obtained from the
observations of  N\,70 by Rosado et al. (1981). 

Models M3 and M4 correspond to model M2 until
t=1.15$\times$10$^5$~yr, at which point we inject a SN (centered for
M3, off-center for M4). Figure~5 shows 
the distributions of density, temperature and radial velocity as function 
of radius for model M3. From the density profile we obtain a shell 
position  between 43 and 52~pc from the center, with a peak density
around R=47~pc. 
The temperature is adequate for X-ray emission inside a region 
of 41~pc in radius. The radial velocity profile show
an average value in the shell around $\sim$62~km~s$^{-1}$. This
velocity is close to the observed value.

\begin{figure}
\includegraphics[width=7.5cm]{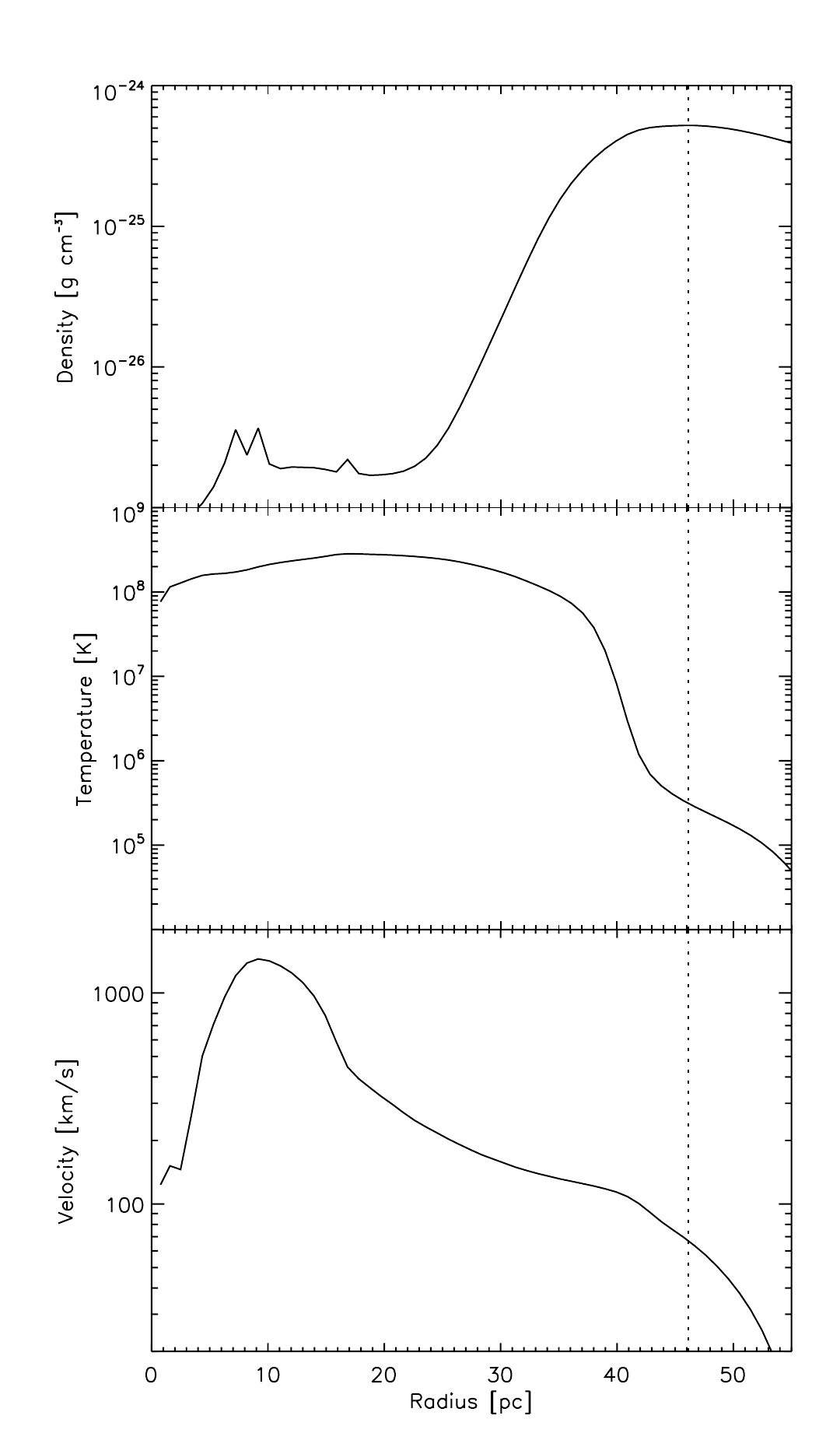}
\caption{Same as Figure~2 but for model M3.}
\end{figure}

Since in M4 the SN not centered one can not assume radial symmetry and
radial averages (eqs.~\ref{ra}-\ref{ta} ) are not longer
appropriate. 
However, in order to estimate an average radial velocity of the shell
in this model, we used the 
equation~\ref{va} and the average radial velocity in shell is 
$\sim$66~km~s$^{-1}$ (see the velocity profile of this model in 
Figure~6), similar to that of M3 model, and also
similar to N\,70 observations.

\begin{figure}
\includegraphics[width=7.5cm]{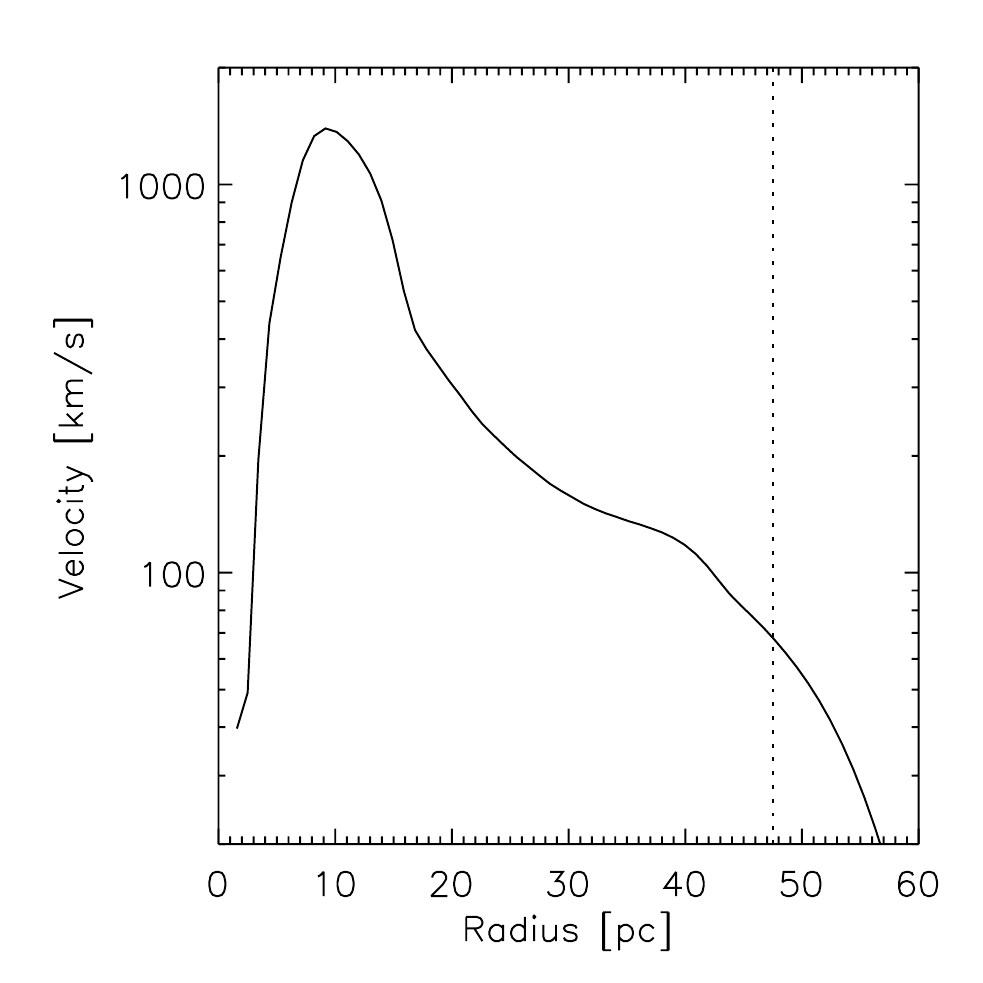}
\caption{The spherically averaged radial velocity flow from model M4,
obtained from the numerical simulation as a function of spherical radius 
$R$.  The dashed lines represent the position of the maximum value of 
the shell density.}
\end{figure}

\section{H$\alpha$ and X-ray emission}

From the results of the simulations we computed  H$\alpha$ 
maps, integrating the emission coefficient along the $x$-axis. The
emission coefficient  is obtained with the interpolation formula given
by Aller (1987) for the temperature dependence of the recombination
cascade.

We also made X-ray emission maps, using  the density and temperature 
distributions from the simulations and plugging them into the CHIANTI
atomic data base and software (see Dere et al. 1997, Landi et
al. 2006).The maps are obtained integrating the X-ray emission
   coefficient along the $z$-axis. For this calculation, it is
assumed  that the  ionization state of the gas corresponds to coronal
ionization equilibrium in the low density regime  (i.~e. the emission
coefficient is proportional to the square of the density). The
emission has been separated into three energy bands  [$0.2- 2$], [$2 - 10$], 
and [$10 - 20$]~keV. The emission coefficient for this energy bands as a
function of temperature is presented in
Figure~7.

\begin{figure}
\includegraphics[width=\columnwidth]{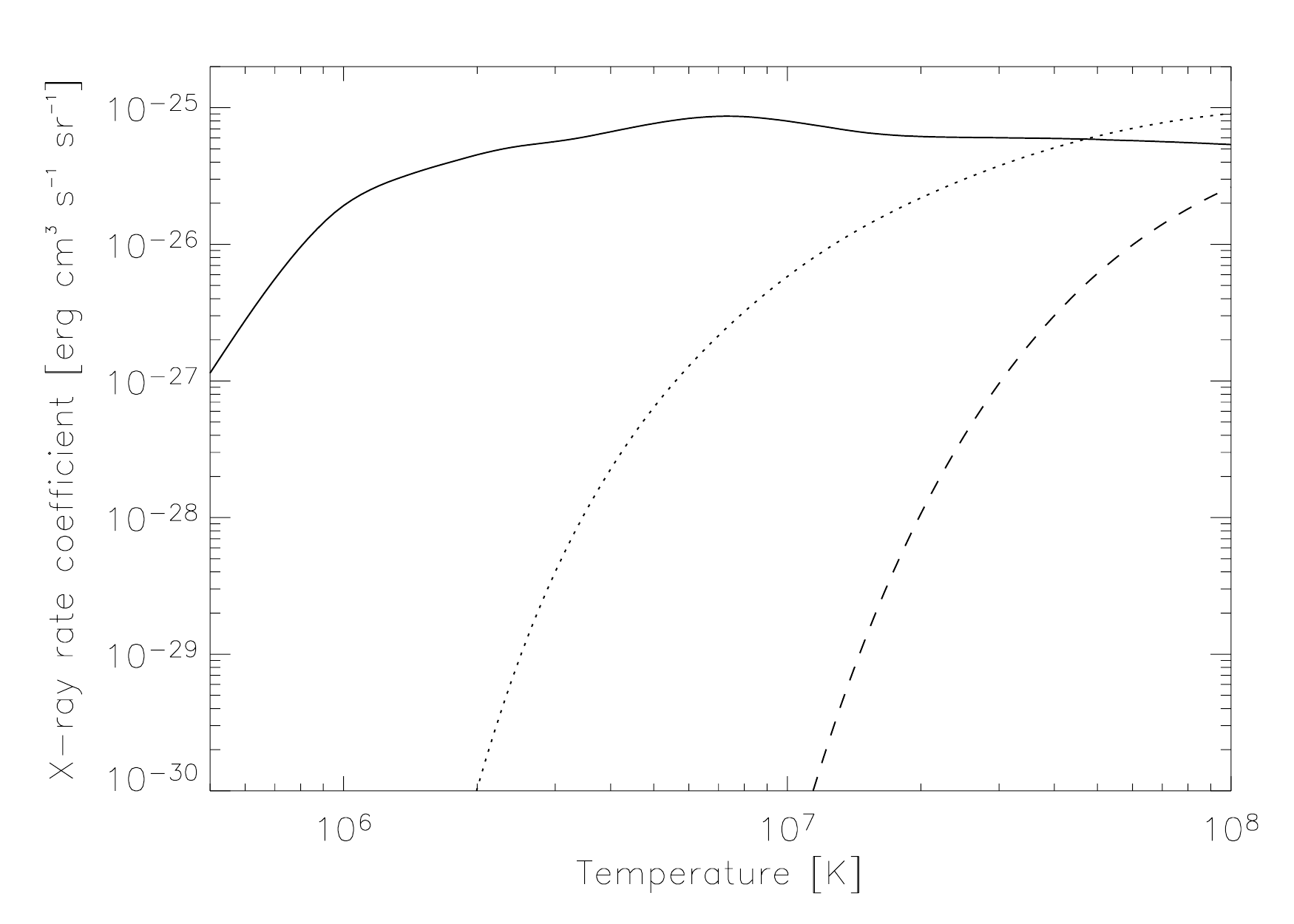}
\caption{X-ray emission coefficients in the [$0.2-2$], [$2 - 10$], 
[$10 - 20$]~keV energy ranges (solid, dotted and
dashed lines respectively) as function of the temperature.}
\end{figure}

We also calculated the X-ray emision for all the models as
function of time. All our models cover a evolutionary time of
2$\times$10$^5$~yr, corresponding approximately to the dynamical age of the 
superbubble derived by Rosado et al. (1981). 
In Figure~8 we present the  X-ray luminosity, in the energy 
range of 0.2 to 2~keV for M1, M2, M3 and M4. 
For visual purposes the horizontal axis of M1 (where the SN was
initiated at $t=0$) was shifted to coincide with the SN starting point
of models M3 and M4 (t=1.15$\times$10$^5$~yr).
From the figure one can see
that the X-ray luminosity for model M2 has a maximum value
of L$_X\sim$4$\times$10$^{34}$~erg~s$^{-1}$ 
(5 times less energy as observed) reached at
t=5$\times$10$^4$~yr. After this time the luminosity slowly declines. 

\begin{figure}
\includegraphics[width=8.9cm]{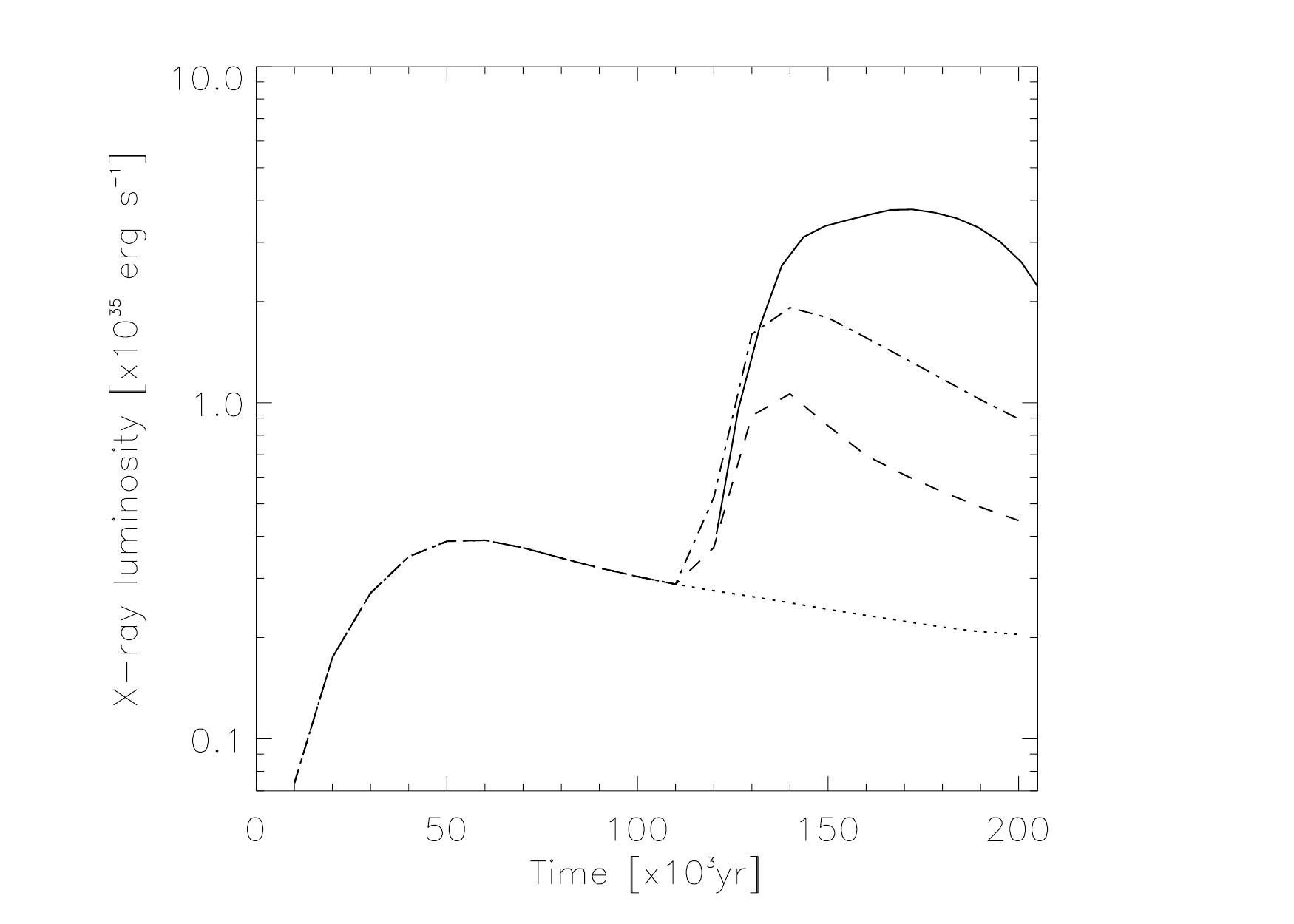}
\caption{X-ray luminosity (between 0.2 to 2~keV) as function of 
time, for models M1, M2, M3 and M4 (solid, dotted, dashed and dot-dashed 
lines respectively. The calculated X-ray luminosities of model M1 was 
shifted by $\Delta t=$1.15$\times$10$^5$~yr, in order to compare the its 
X-ray luminosity values with the X-ray luminosities of M3 and M4 
after the supernova explosion}
\end{figure}

The rest of the models, in which we have included a SN, reach X-ray  
luminosities of $>10^{35}$~erg~s$^{-1}$ (see Table~3). In model M1, the 
highest value of the X-ray luminosity is
$\sim$3$\times$10$^{35}$~erg~s$^{-1}$, and this luminosity is kept at
or above the N\,70 observed value for $\sim$5$\times$10$^4$~yr. However, 
when this model reaches the observed radius value of N\,70 ($\sim$50~pc,
at t=2$\times$10$^5$~yr), the X-ray luminosity has dropped by more
than 2 orders of magnitude below the observed value. 

\begin{deluxetable*}{cccccc}
\tablecaption{Models results}
\tablewidth{0pt}
\tablehead{
\colhead{}&\colhead{$v_s$} &\colhead{L$_{X,rs}$}&\colhead{t$_{\rm Lx}$}&
\colhead{$\Delta$t$_{\rm Lx}$}&\colhead{L$_{X,max}$}\\
\colhead{}&\colhead{[km s$^{-1}$]}&\colhead{[10$^{35}$ erg s$^{-1}$]}&
\colhead{[10$^5$~yr]}&\colhead{[10$^3$~yr]}&\colhead{[10$^{35}$ erg s$^{-1}$]}
}
\startdata
Obs.           &  70\tablenotemark{a} & 1.6\tablenotemark{b}   &2.4 &  --  & --\\
M1             &  90     & $\sim$ 2.0 & 2.7\tablenotemark{*}& 52 & 3.76  \\ 
M2             &$\sim 45$& $\sim$ 0.21&  --    &  -- & 0.364 \\
M3             &$\sim 67$& $\sim$ 0.48 &  1.20  & 2.0 & 1.02\\
M4             &$\sim 75$& $\sim$ 1.00&  1.18  & 75  & 1.96\\ 
\enddata
\tablenotetext{a}{Rosado et al. (1981)}
\tablenotetext{b}{Reyes-Iturbide et al. (2011)}
\tablenotetext{*}{The calculate X-ray luminosities of model M1 was 
shifted by $\Delta$t=1.15$\times$10$^5$~yr}
\tablenotetext{}{Where, L$_{X,rs}$ is the X-ray luminosity at the time 
the model reaches a size similar to that of N\,70, t$_{\rm Lx}$ is the 
evolutionary time when the maximum X-ray luminosity is reached for each 
model and $\Delta$t$_{\rm Lx}$ is the interval that the X-ray luminosity is 
kept above 10$^{35}$~erg~s$^{-1}$}
\end{deluxetable*}

 The maximum X-ray emission in model M3 can reach
$10^{35}$~erg~s$^{-1}$, but it is still significantly lower than the
observed luminosity, and when the superbubble reaches the observed 
radius the X-ray luminosity is already 5 times smaller. A
centered supernova explosion at t=10$^5$~yr (when the shell is closer to the 
center of the stellar cluster) could help to reach the N\,70 X-ray emission,
but by the time it reaches a 50~pc radius the luminosity would be
down to a value of $\sim 3\times$10$^{34}$~erg s$^{-1}$ (comparable to 
model M2).

The maximum X-ray luminosity, in model M4, is 
$\sim$2$\times$10$^{35}$~erg~s$^{-1}$, and this luminosity is above the 
observed value for a timescale of $\sim$7.5$\times$10$^4$~yr. By the time 
the superbubble reaches a radius of $\sim$50~pc the X-ray luminosity
agrees well with the observations.

In Figure~9 we show synthetic X-ray emission maps for
  model M2. The emission in the figure has been separated into three
  energy bands $0.2-2$, $2-10$, and $10-20$~keV (from top to bottom,
  panels (a), (b), and (c), respectively). 
It is readily evident that the emission is dominated by soft X-rays
with only a small contribution from harder X-rays. For this model (M2) the
emission in the soft X-ray band ($0.2-2$ keV) is three orders of
magnitude larger than the emission in the $2-10$~keV energy range
and over five orders of magnitude larger than the harder
X-ray emission ($10-20$~keV). 
The fact that the emission in hard X-rays negligible with respect to that
in soft X-rays might seem surprising at a first glance, considering
that there is a large region (inner $30$ pc) filled with  10$^8$~K
gas, which should emit in hard X-rays (see the emission coefficients
in Figure~7). However, the density at the interior of the bubble is quite
low, and it is only beyond $\sim 30$ pc that it increases (rapidly)
with radius, at the same temperature drops to ${\sim
  10^7~\mathrm{K}}$. 
Since the thermal  X-ray emission is proportional to the density
squared the result is that most of the emission observed arises from
close to the shell, form a region cold enough to produce soft X-rays.

Temperatures of 10$^8$~K have been observed and modeled in super stellar
clusters (Silich et al. 2004 and 2005), which are much more massive that the 
young star association in N\,70. The reason for such temperatures is the
high terminal velocity of some of the winds
($>2000~\mathrm{km~s^{-1}}$). The difference is that in super stellar
clusters the density of stars is  significantly larger, thus the
gas density inside is enough to produce an observable amount of hard
X-rays. In contrast, the massive stars in N\,70 are too far apart each other
and
the emission above $2~\mathrm{keV}$ is very faint compared with that 
at lower energies.

\begin{figure}
\includegraphics[width=8.9cm]{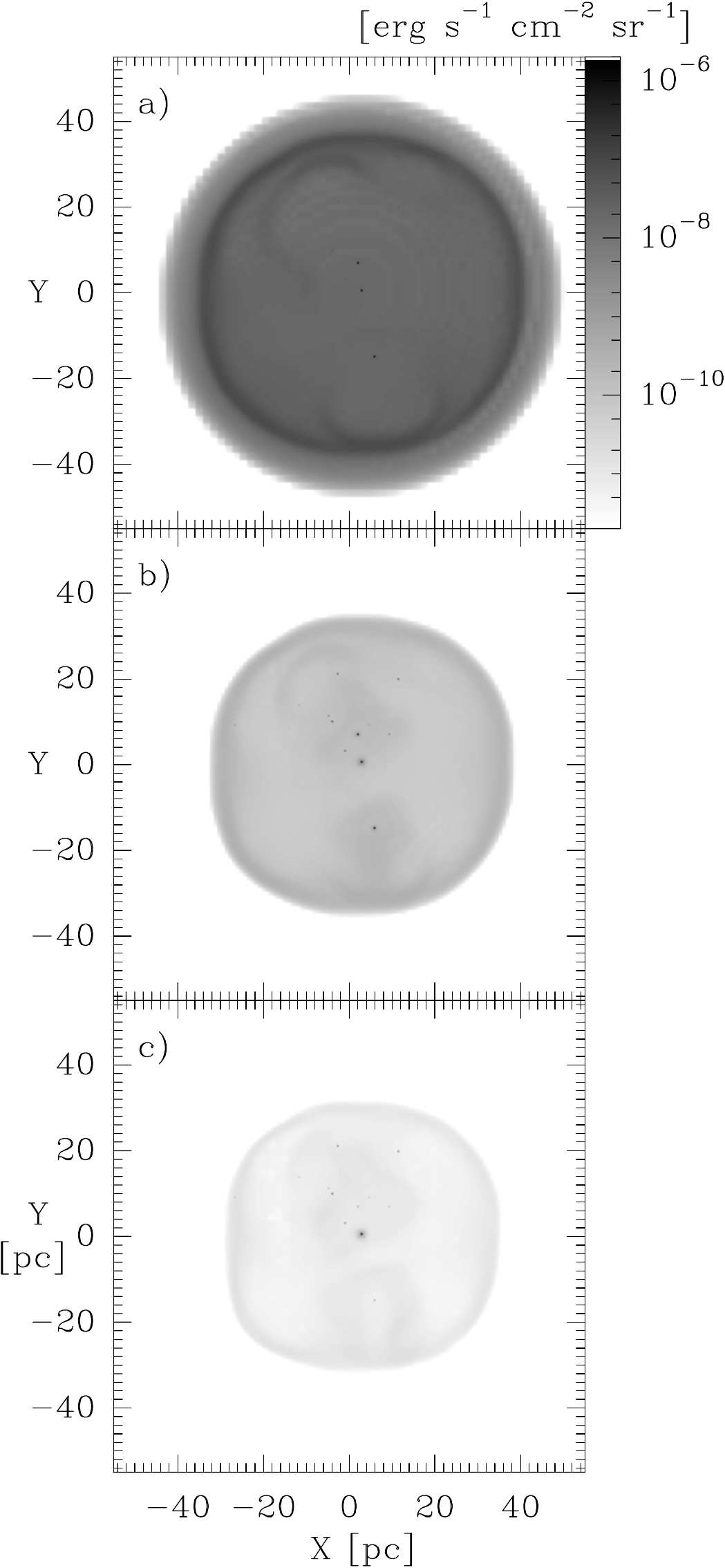}
\caption{Synthetic X-ray emission map of model M2 in
    the (a) [$0.2-2$]~keV, (b) [$2-10$]~keV, and (c) [$10-20$]~keV 
energy ranges.}
\end{figure}

Figure~10 shows the H$\alpha$ map and superposed X-ray isocontours of M4 
model. The X-rays isocontours cover a wide range of  flux of energy from 
10$^{-9}$ to 10$^{-6}$~erg~cm$^{-2}$~s$^{-1}$~sr$^{-1}$ 
in steps of
5$\times$10$^{-8}$~erg~cm$^{-2}$~s$^{-1}$~sr$^{-1}$. The highest
values of the isocontours are at the center and in a shell just behind
(inside) the optical superbubble (between 38 to 47~pc). The outer
shell or superbubble is formed by the interaction of the cluster 
wind and its surrounding ISM.

\begin{figure}
\begin{center}
\includegraphics[width=7.5cm,angle=-90]{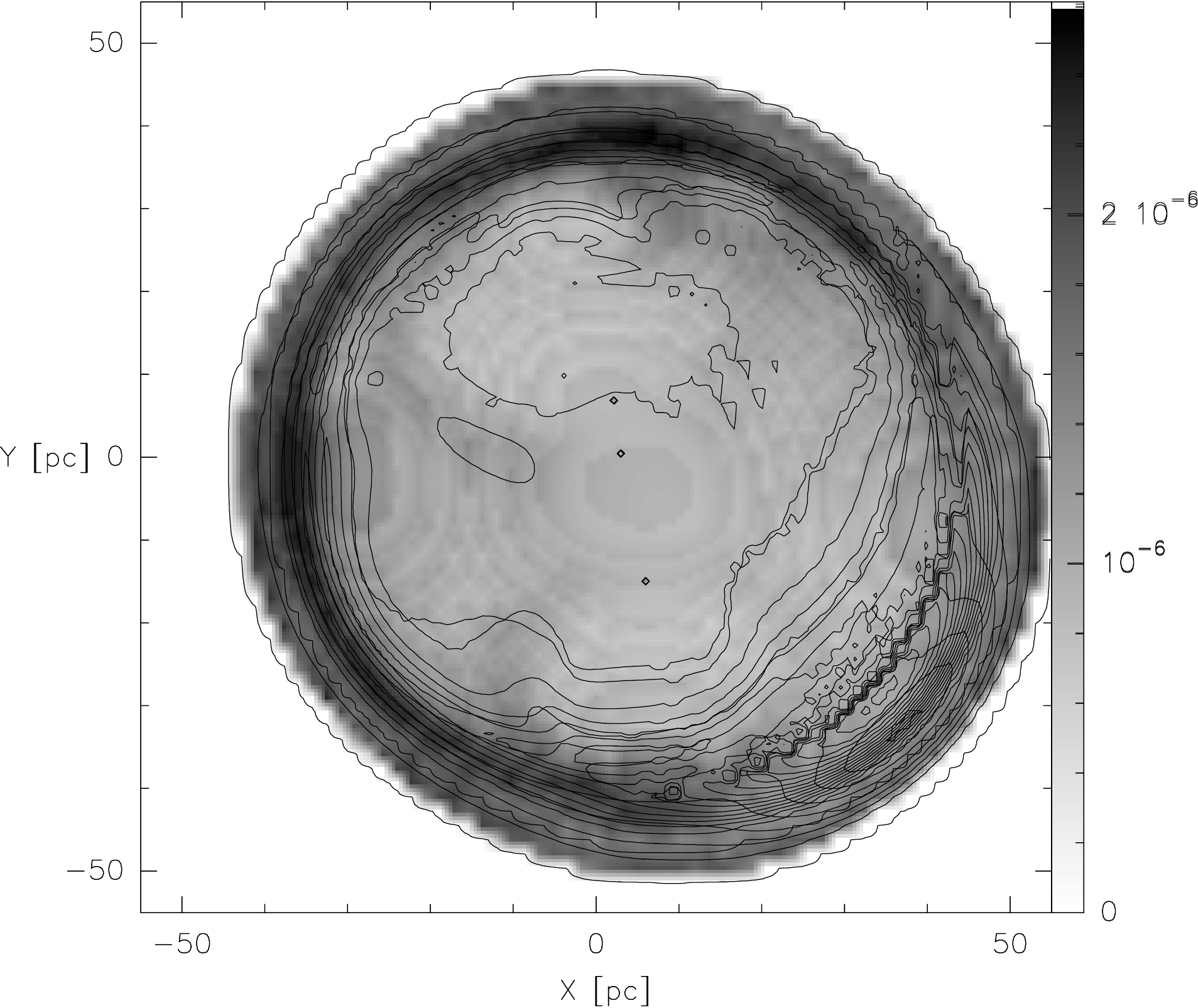}
\caption{Overlay of the simulated H$\alpha$ emission map (grays) with contours of
the synthetic X-ray emission.}
\end{center}
\end{figure}

Table~3 present a summary of the numerical results for the dynamics 
and X-ray luminosities obtained from our models. In this table we include
the evolutionary time when the maximum X-ray luminosity is reached 
for each model ($t_{Lx}$) and the interval that the X-ray luminosity is
kept above 10$^{35}$~erg~s$^{-1}$ ($\Delta$t$_{\rm Lx}$).

In our models  we did not include the thermal conduction effects. Weaver
et al. (1977) and several other authors (Chu \& Mac Low 1995, Silich et al. 2001
etc.) have recently studied its
importance to explain the total X-ray emission in stellar clusters 
and SNRs. However, Silich et al. (2001) shows that while thermal
conduction might have produce an enhancement of several orders of
magnitude in superbubbles with ages  $>$10~Myr, for young superbubbles
(such as N\,70) thermal conduction can only produce a difference of a
factor of $\sim5$. It is important to notice that the main effect of thermal 
conduction is to carry material from the external shell into the bubble, thus
filling the bubble with X-rays, and maintaining its emission for a
longer time (see also Silich et al. 2001). This is because thermal conduction
drives a transfer of the material from the external shell to the center of
the bubble.

The standard model of bubbles (Weaver et al. 1977) predicts X-ray emission 
from the hot interior of bubbles by including thermal conduction effects. 
Its success is controversial because in some cases the predicted X-ray 
luminosities where lower than detected (as in the case of the N70
superbubble) 
while, in other cases, the predicted X-ray emission is higher than detected 
(as in the case of the M17 superbubble; Dunne et al. 2003, Reyes-Iturbide 
et al. 2009). The new results on thermal conduction effects mentioned above 
make us believe that thermal conduction is not the main ingredient 
originating the difference. In this work we propose that the inclusion
of a supernova explosion, as an additional agent to be  
considered besides the stellar winds, is more important than thermal
conduction. 
At least two reasons could be given in order to support this: (1) In the case 
of M17, it is almost certain that no SN explosion has occurred yet while the 
age of LH114, at the interior of N70, makes plausible a SN explosion,
and (2) the expansion velocities predicted by Weaver et al. (1977)
model in the case of  
N70 are much lower than the measured velocities for this superbubble. As seen 
in Figures 2 to 5 and Table 3 only the models including a SN explosion predict 
a shell acceleration that could explain large expansion velocities as the 
ones measured in high-velocity shells, such as N70. Thus, we suggest
that the main  
difference between high velocity and low velocity superbubbles is the 
occurrence (or lack) of a SN explosion in their interiors. Off-centered
explosions can change some of the detailed structure and dynamics, but
the main conclusions remain unchanged.
Of course, we have explored only the N70 superbubble and we need to
study in detail other superbubbles (both of high and low-velocity
types)in order to confirm this suggestion.

\begin{figure}
\begin{center}
\includegraphics[width=7.5cm]{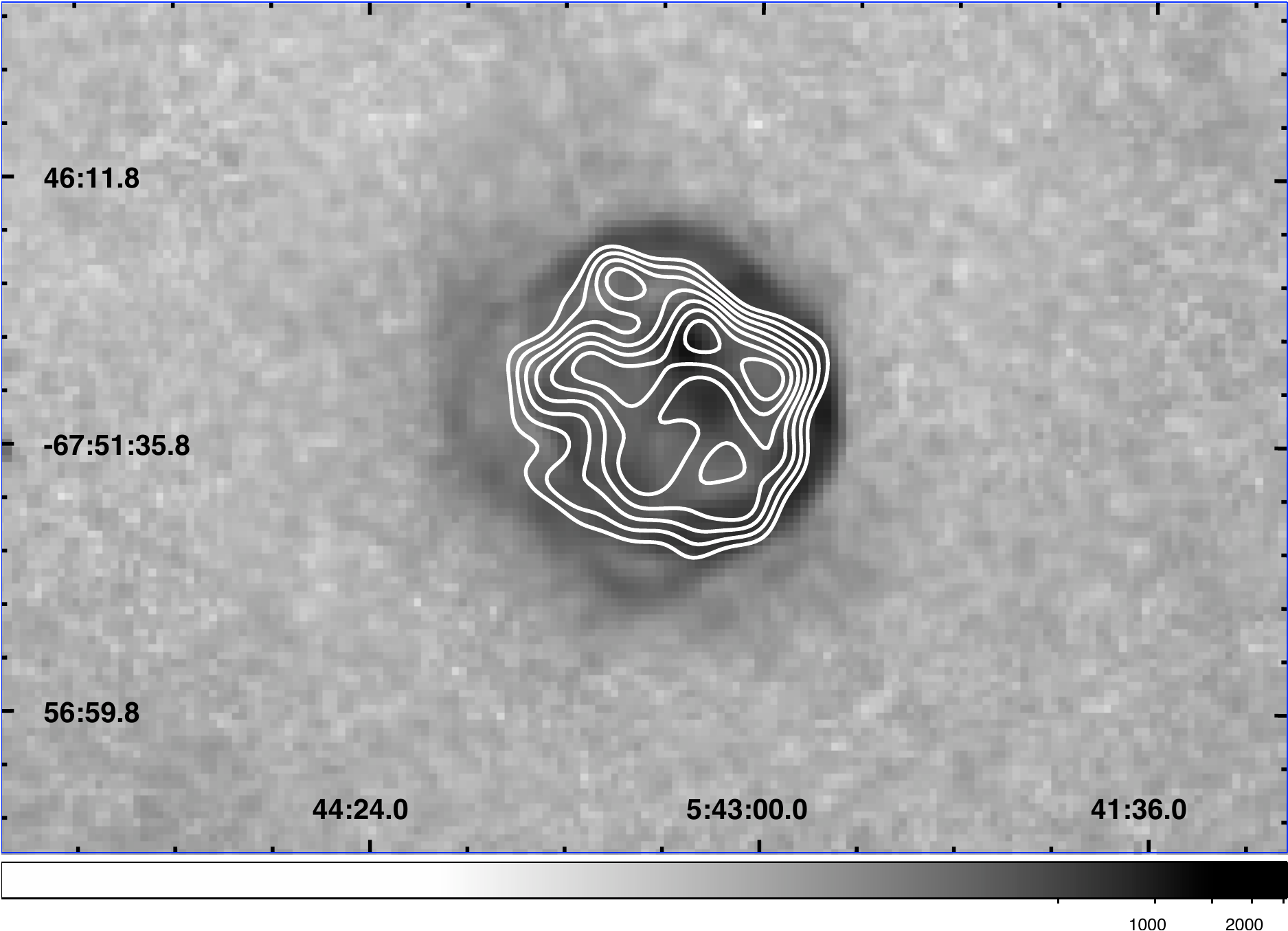}
\caption{The Halpha MCELS image of N70 (grays) is overlaid with X-ray contours.
Contours have been drawn at levels of 3 , 4 , 5 , 6 , 7 , 8 , 9  and 10  
above the background level.}
\end{center}
\end{figure}

\section{Conclusion}

We have studied the dynamics and X-ray emission of superbubbles
  driven by cluster winds including in our models supernova explosions
  alone, stellar winds alone, and a combination of stellar winds and
  SN explosions, these latter at different times and locations.  We have
  turned our attention to the superbubble N\,70 in order to confront
  our model predictions with the observations.  We computed four
  models (M1-M4) of superbubbles using the properties of the more
  massive stars contained in the cluster inside the N70 superbubble,
  adopting the ISM density and metallicity around this
  superbubble. The models are evolved in a homogeneous (in density and
  temperature) medium.

From our models we demonstrated that the case in which only the
stellar winds inject mechanical energy (M2), the soft X-ray luminosity
is lower by an order of magnitude than the observed value (in
agreement with the standard model). And the radial velocity of the
shell is less than 45~km~s$^{-1}$.  However, the model of a single
supernova explosion (M1), even when the input from stellar winds is
not considered, could reach the X-ray luminosity and an expansion
velocity consistent with the observations. Nevertheless, a single SN
explosion predicts the formation of a very thin shell which is not in
agreement with the morphology of the N\,70 superbubble.

Three models considered the mechanical energy injected by stellar
winds, M2 only considers the input from stellar winds, while M3 and M4
have been combined with a SN explosion. 
We included the SN explosion at two different positions,
near to the cluster center, and $\sim$2~pc from the cluster
center (M3 and M4 respectively). The SN has exploded after
t=1.15$\times$10$^{5}$~yr of the 
evolutionary time of the cluster wind.  From models M3 and M4 we can
obtain an X-ray emission in good agreement with the observational
data during 20 and 75~kyr, respectively.  And the shell velocity
expansion ($\sim$60~km~s$^{-1}$), obtained in both models, could
explain the kinematics measured for this bubble. Models M3 and M4
formed a thick shell, also in agreement with the observations of
N\,70.

As a matter of fact, both models M3 and M4 reproduce quite well  the large 
measured expansion velocity of the N70 shell and the X-ray luminosity. 
Model M4 lacks spherical symmetry because the off-centered SN, however,
the morphological difference is somewhat subtle 
and it can be concealed for certain orientations with respect
to the line of sight. So, we cannot discard it. Figure 10 shows the
 predicted H$\alpha$ emission (gray levels) and the X-ray emission 
(isocontours) for the M4 model and  Figure~11 depicts the observed 
ones showing good agreement. 

It is important to notice that our models predict a large region
inside the superbubble (the innermost $\sim 30~\mathrm{pc}$) with
temperatures  $\gtrsim 10^8~\mathrm{K}$, which would result in thermal hard
X-ray emission (above $2~\mathrm{keV}$). However the density inside
the superbubble is very low and it produces only a faint emission that
is overwhelmed by the soft X-rays produced in the surrounding shell.

We end by noting that in this paper we  did not include the thermal 
conduction effects.  However, for young superbubbles, 
with ages less than 10~Myr (as well as N\,70) the differences  between
models with and without thermal conduction are only on a factor of
$\sim 5$ in $L_X$ (Silich et al. 2001).
A more important role of thermal conduction in superbubble models is
the fact that it helps sustain the X-ray emission for longer periods
of time. 
\acknowledgments
Invaluable comments of Sergiy Silich are deeply appreciated. A.R.-G.
is grateful with the hospitality of the INAOE. This paper recived
financial support from grants 40095-F (CONACYT), IN102309 and IN119709
(DGAPA-UNAM).

\end{document}